\documentclass[twocolumn,showpacs,amsmath,amssymb,pra,superscriptaddress,floatfix,raggedfooter]{revtex4}

\usepackage{amsmath}
\usepackage{amssymb}
\usepackage{graphicx}
\usepackage{epsfig}
\usepackage{amsfonts}
\usepackage{hyperref}
\usepackage{color}

\begin{document}

\title{Spontaneous Peierls dimerization and emergent
bond order in one-dimensional dipolar gases}

\author{M. Di Dio}
\affiliation{IOM-CNR Democritos, SISSA Via Bonomea 265, I-34136 Trieste, Italy}

\author{L. Barbiero}
\affiliation{Laboratoire de Physique Th\'eorique, CNRS, UMR 5152 and Universit\'e de Toulouse, UPS, F-31062 Toulouse, France}
\affiliation{Dipartimento di Fisica e Astronomia "Galileo Galilei", Universit\`a di Padova, I-35131 Padova, Italy}

\author{A. Recati}
\affiliation{INO-CNR BEC Center, via Sommarive, 14, I-38121, Povo, Trento, Italy}

\author{M. Dalmonte}
\affiliation{Institute for Quantum Optics and Quantum Information, and Institute for Theoretical Physics,
  Technikerstra\ss e 21a, A-6020 Innsbruck, Austria}

\begin{abstract}
We investigate the effect of dipolar interactions in one-dimensional systems
in connection with the possibility of observing exotic many-body effects with trapped atomic and molecular
dipolar gases. By combining analytical and numerical methods, we show how the competition between 
short- and long-range interactions gives rise to frustrating effects which lead to the stabilization
of spontaneously dimerized phases characterized by a bond-ordering. This genuine quantum order is sharply distinguished from Mott and spin-density-wave phases, and can be 
unambiguously probed by measuring non local order parameters  via {\it in-situ} imaging techniques.
\end{abstract}


\maketitle

\section{Introduction}
Cold atom gases confined in reduced dimensionality represent an 
ideal system to observe many-body phenomena related
to the prominent role played by quantum fluctuations~\cite{bloch_review,cazalilla2011}.
Recent experimental
advances have paved the way to the investigation of quantum magnetism,
notable examples being the demonstration of super-exchange interactions
in bosonic gases~\cite{anderlini2007}, the time evolution of spin impurities~\cite{minardi2012,bloch2013}, 
the observation of
frustrated classical dynamics~\cite{sengstock2012} and the engineering of 
Ising~\cite{greiner2011} and anisotropic exchange Hamiltonians~\cite{esslinger2013}. 

New opportunities in this direction are now stimulated by the prominent
progress in cooling and controlling ultracold gases of magnetic atoms 
and polar molecules, which provide tunable platforms where the effect of long-ranged  
dipolar interactions can play a dominant role in determining the many-body
dynamics~\cite{PolMolExp,lahaye2007,pasquiou2010,lu2011,aikawa2012}. Such progress has opened a new door for the investigation of 
lattice models beyond the conventional Hubbard paradigm, where long-range
interactions can compete with local ones on the way to unveil richer many-body
physics~\cite{carr_review,LewensteinReview, baranov_review}. 
Much attention has been devoted up to now to extended Bose-Hubbard 
models, where a new phase of matter, the Haldane insulator (close analog of the 
Haldane phase in spin-1 chains) has been predicted to occur~\cite{dallatorre2006,
dalmonte2011,santos2012,Rossini2012}. However, not much 
is known on other possible magnetic phases for fermionic dipolar gases in optical
lattices~\cite{baranov_review}, which present close analogy to the
so called extended Hubbard model (EHM)~\cite{bari1971,hirsch1984,cannon1990,nakamura2000,sengupta2002,furusaki2002,
jeckelmann2002,sandvik2004,furusaki2004,tam2006,ejima2007}.

In this  article we show that one-dimensional (1D) Hubbard models with long-range interactions support
a non-trivial insulating phase characterized by bond order~\cite{nakamura2000,ejima2007,sandvik2004,furusaki2002} [a bond-order density wave (BOW)]
due to the competition between dipolar and on-site interactions. 
This phase has attracted notable interest in recent years in the context
of strongly correlated electron systems. Its existence is now well established but it has been long debated in a series of theoretical studies~\cite{nakamura2000,sengupta2002,furusaki2002,
jeckelmann2002,sandvik2004,furusaki2004,tam2006,ejima2007}. The opportunity of realizing and 
observing such states of matter may shed new light on a series of issues, from its
dynamical properties to its ground state robustness. In order to prove its
existence in dipolar Hubbard models, we combined analytical and state-of-the-art
numerical methods based on the density-matrix renormalization group (DMRG)
algorithm~\cite{white1992,schollwock2005}. Remarkably, the effect of dipolar interaction is evident
already in the weak-coupling approach, leading to the prediction of a BOW phase
within one-loop order. This is in sharp contrast with standard EHMs - on-site and nearest neighbor interaction only-, where a simple
formalism is instead unable to capture its existence~\cite{nakamura2000}.
While the detection of the BOW phase is 
in general challenging due to the limited extension in parameter space
and its lack of density-like order parameters, we show how the recently
developed {\it in situ} imaging techniques~\cite{bakr2009,endres2011} provide an ideal route toward
the unbiased identification of such phases since the BOW
phase is uniquely identified by the long-range order of non local
parity correlations~\cite{montorsi2012,barbiero2013}.

\begin{figure}[tb]
\begin{center}
\includegraphics[height=53mm]{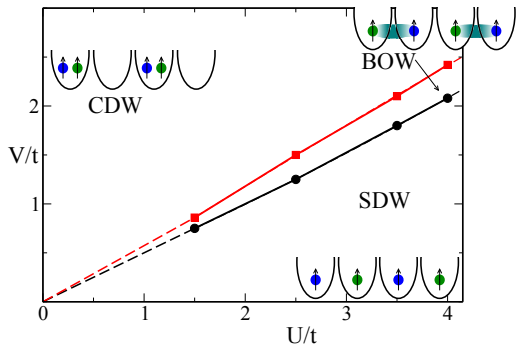}
\caption{(Color online) Numerical phase diagram for dipolar mixtures described by the 
Hubbard Hamiltonian in Eq.~\eqref{H_dis}. In between a CDW and a SDW phase, a region with dominant bond-order develops
close to the line $U\simeq V\frac{3\zeta(3)}{2}$; the CDW-BOW and BOW-SDW transitions are of the Gaussian and Berezinskij-Kosterlitz-Thouless type respectively. Errors in determining phase transition points are smaller than symbol sizes. The arrows in the cartoons denote the alignment of the dipoles, while green  (light gray) and blue (dark gray) points represent the internal spin.}\label{fig1}
\end{center}
\end{figure}

\section{Spin-1/2 dipolar Fermi gas}

Dipolar particles confined in a one dimensional tube can be described by the 
following microscopic Hamiltonian:
\begin{eqnarray}\label{Hdip}
&H&= \int \;dx \sum_{\sigma=1,2}\left\{\psi_\sigma^\dagger(x) \left[-\frac{\hbar^2}{2m}\partial_x^2\right] \psi_\sigma(x)\right\}\\
&+&\sum_{\sigma, \sigma'}\int dx \;dy \big[\rho_\sigma(x)(\mathcal{V}(x,y)+\mathcal{U}\delta(x,y)) \rho_{\sigma'}(y)\big],\nonumber
\end{eqnarray}
where $m$ is the particle mass, $\psi^\dagger_\sigma(\psi_\sigma)$ are creation~(annihilation) operators of
$\sigma=1,2$ fermions, and $\rho_\sigma=\psi^\dagger_\sigma\psi_\sigma$. The short-distance contribution of interspecies interactions $\mathcal{U}$ can be 
tuned by using Feshbach resonances or confinement induced, while the dipolar interaction
$\mathcal{V}$ is controlled by means of electric (polar molecules) 
or magnetic (magnetic atoms) fields~\cite{baranov_review}. The pseudo-spin degree of freedom can either be represented by different $m_F$ states (for atoms), or by preparing molecules in different nuclear or rotor states~\cite{gorshkov2011}. Once confined on a
sufficiently deep optical lattice, an effective description in terms of the Hubbard 
Hamiltonian leads to the discrete formulation
\begin{equation}\label{H_dis}
H=H_1+H_2+H_{12} 
\end{equation}
(see, e.g., \cite{bloch_review}), where the single species contributions read:
\begin{equation}\label{H_sigma}
H_{\sigma}=-\sum_{i}t(c^{\dagger}_{\sigma,i}c_{\sigma, i+1}+\textrm{H.c.})+
\sum_{i<j} V (i-j) n_{\sigma,i}n_{\sigma,j},
\end{equation}
with the first and second term describing tunneling and dipolar interactions respectively, and the interspecies coupling is
\begin{eqnarray}\label{H_tb}
H_{12}=U\sum_i n_{1,i}n_{2,i}+ \sum_{i\neq j}V (i-j) n_{1,i}n_{2,j},
\end{eqnarray}
where the first term is the on-site interaction $U$ (given by a combination of local and dipolar
potentials), and the last term describes interspecies dipolar interactions. From here on, we will focus on purely repulsive dipoles, $V(i-j)=V/(|i-j|)^3$, as relevant, e.g., for electrically polarized molecular gases.
In the equal mass, equal interaction case, the system inherits a global SU(2) symmetry,
which is preserved by the long-range tail, and is reduced to U(1) for general parameter
choices. From here on, we will focus on the former case and consider a balanced half-filled chain, 
$N_1=N_2=N/2=L/2$, where $N_\sigma$ is the number of particles in the spin state $\sigma$ and $L$ the system size.
Like in the case of the EHM, the phase diagram of Eq.~(\ref{H_dis})
is determined by the competition between local and non-local interactions.
Before searching for specific BOW instabilities, we illustrate the competing
mechanism in the atomic limit $t=0$, as such competition will then lead to 
spin-frustration at the origin of bond-order itself.

For dominant $U$ interactions, the system ground state is a spin density wave (SDW). On the other hand, a dominant dipolar interaction
will minimize the energy by imposing double occupancies every second site, thus
stabilizing a fully gapped charge density wave (CDW). These two ground states, illustrated in the insets
of Fig.~\ref{fig1}, become energetically
degenerate along the line $U_c^{(cl)}=3\zeta(3)V/2$ ($\zeta$ is the Riemann
zeta function), which determines a phase transition between SDW and CDW.
Deeply in the quantum regime $t\simeq V\simeq U$, however, quantum fluctuations 
may enhance the emergent frustration close to the classical transition line and 
lead to a different critical scenario. For the EHM, where only nearest-neighbor
interactions are considered, 
it was argued by Nakamura~\cite{nakamura2000} that an additional
phase with dominant charge bond-order instability occurs between 
the SDW and the CDW phases.
This phase is characterized by a spontaneous spin-Peierls dimerization,
manifest in a charge polarization on alternating bonds and by the 
formation of spin dimers on the bonds, and constitutes a notable 
example of dimerization in strongly correlated systems due to 
frustration. A two-dimensional analog of this phase has been recently discussed
for spinless dipolar fermions in layers~\cite{bhongale2012}.

The existence and extent of the BOW phase in the EHM have been intensively debated mainly because 
(i) the instability is not captured by one-loop \textit{g}-ology calculations
based on bosonization (while more refined methods recently provided
analytical evidence of it~\cite{furusaki2002,tam2006}), and 
(ii) numerical results were not consistent due to the small extent in
parameter space of such a phase, and to the difficulty of providing an 
accurate location of the critical line~\cite{nakamura2000,jeckelmann2002,sandvik2004,ejima2007}. In the following, combining analytical and numerical methods, we show that
dipolar systems support spontaneous spin-Peierls dimerization, and 
that the corresponding BOW phase occupies a larger region in 
parameter space with respect to the EHM. 
First of all, we show that one-loop \textit{g}-ology is sufficient to establish the 
existence of bond-order in the weak-coupling regime. Then, the existence and 
extent of the BOW phase are benchmarked with DMRG simulations.

\section{Low-energy field theory}

We now present a qualitative study of Eq.~\eqref{H_dis} within the bosonization
framework~\cite{bosonization}. As a first step, we express the fermionic lattice operators in terms of 
continuum chiral fields $\psi^\dagger_{R/L}(x)$:
\begin{equation}
c^\dagger_{j,\sigma}=\sqrt{a}[\psi^\dagger_{R,\sigma}(x)e^{ik_F x}+\psi^\dagger_{L,\sigma}(x)e^{-ik_F  x}],
\end{equation}
with $x=j a$, $a$ being the lattice spacing and $k_F=\pi N/(2 L a)$ the Fermi momentum.
We then apply the standard bosonization mapping introducing 
density and phase fluctuation fields $\varphi_{\sigma},\vartheta_\sigma$ 
for the two species in order to map the original fermionic
problem onto a bosonic one:
\begin{equation}
\psi_{R,\sigma}=\frac{\eta_{R,\sigma}}{\sqrt{2\pi a}}e^{-i[\varphi_{\sigma}-\vartheta_{\sigma}]}, \quad \psi_{L,\sigma}=\frac{\eta_{L,\sigma}}{\sqrt{2\pi a}}e^{i[\varphi_{\sigma}+\vartheta_{\sigma}]}
\end{equation}
where $\eta_{r,\sigma}$ are the so-called Klein factors. 
Typically, the bosonization mapping is applied to the microscopic Hamiltonian, taking as a starting point non-interacting fermions, and thus deriving the effective parameters in the low-energy theory in a perturbative manner, limiting the validity to the regime $V, U\ll t$. Here we take advantage of recent analytical and numerical results, and explore an alternative route to derive the coefficients in the effective low-energy theory, typical, e.g., in the Landau theory of Fermi liquids \cite{FermiLiquid}. Our starting point is the single species Hamiltonians Eq. (\ref{H_sigma}), and not the free tunneling Hamiltonians as usual. In their gapless regime, i.e. when $V/t\lesssim2\zeta(3)$, their bosonized form reads
\begin{eqnarray}
H_\sigma&=&\frac{\hbar v}{2}\int dx \left[\frac{(\partial_x\varphi_\sigma)^2}{K}+
K(\partial_x\vartheta_\sigma)^2 \right]. 
\end{eqnarray}
The single-species Luttinger parameter is well approximated by the continuum theory estimate $K=(1+1.46n*V/t)^{-1/2}$~\cite{citro2007, citro2008, roscilde2010,dalmonte2010}. While this result strictly holds in the continuum, its quantitative behavior has been verified also in lattice calculations~\cite{dalmonte2011b}, as far as the crystalline instability is not approached. This ensures a controlled estimate of $K$ in the regime $V/t\lesssim 1$, and possibly beyond. We notice that embodying interactions in a non-perturbative (although approximate) manner in bosonized Hamiltonians is an established procedure, which allows one to retain features usually missing in perturbative treatments (see, e.g., \cite{bosonization}).

We then proceed and, on the top of the interacting single species Hamiltonian, we consider the role of $H_{12}$. We first introduce density and spin collective fields:
\begin{equation}
\varphi_{c/s}=\frac{\varphi_1\pm\varphi_2}{\sqrt{2}}, \quad
\vartheta_{c/s}=\frac{\vartheta_1\pm\vartheta_2}{\sqrt{2}}.
\end{equation}
After bosonizing the interspecies interaction $H_{12}$, the Hamiltonian $H$ can be decoupled into a density and spin sector, $H=H_c+H_s$, where the physics in each sector is described by a sine-Gordon model:
\begin{eqnarray}
H_{c/s}&=&\frac{\hbar v_{c/s}}{2}\int dx \left[\frac{(\partial_x\varphi_{c/s})^2}{K_{c/s}}+
K_{c/s}(\partial_x\vartheta_{c/s})^2 \right] \nonumber\\
&+& g_{c/s}\int dx \cos[\sqrt{8\pi}\varphi_{c/s}].
\end{eqnarray}
Notice that terms coupling density and spin degrees of freedom are also
present; in the following we will neglect their effects, which are supposed
to be small away from the regime $V, U\ll t$ due to their
large scaling dimension. The coefficients in $H_{c/s}$ can be extracted treating $H_{12}$
as a perturbation on the top of the decoupled $H_1$ and $H_2$; here, the local interactions are treated as in the conventional Hubbard model~\cite{bosonization}. In the spin sector, we compute the effect of the interspecies interactions on the quadratic part of the 
Hamiltonian, and then fix the coefficient of the mass term by imposing SU(2) symmetry - which has to be retained exactly at low-energies.  A similar procedure is illustrated in Ref.~\cite{bosonization} in the context of the EHM model: here, the SU(2) symmetry is recovered by fixing $g_s = 1-1/K_s$.
The sine-Gordon model then supports a Berezinskij-Kosterlitz-Thouless 
(BKT) transition~\cite{bosonization} at $K_s=1$, that, from its microscopic form, is 
\begin{equation}
K_{s}=\sqrt{K\left(\frac{1}{K}-\frac{U+2\zeta(3)V}{2\pi t}\right)^{-1}}.
\end{equation}
At weak coupling, the transition occurs at $V_c^{(BKT)}\simeq U/2.18$.
In the charge sector, the Luttinger parameter $K_c=\sqrt{K/\left(\frac{1}{K}+\frac{U+2\zeta(3)V}{2\pi t}\right)}$ always satisfies $K_c<1$ (since $K<1$): as such, phase transitions in this sector only depend on the coefficient $g_c$. In particular, $g_c=0$ defines a Gaussian phase transition line between phases where the cosine potential in $H_c$ is pinned at different values (at strong coupling, this line can
be unstable toward $4k_F$ mass terms when $K_c<1/4$). The perturbative estimate of $g_c$ reads
\begin{equation}
g_c\propto U-2V\sum_{j=1}^\infty \left[\frac{1}{(2j-1)^3}-\frac{1}{(2j)^3})\right]= U-\frac{3\zeta(3)V}{2}.
\end{equation} 
implying a weak-coupling Gaussian transition at $V_c^{(G)} \simeq U/1.80$.
The phase diagram can be subsequently mapped out by considering the 
dominant orders as in the case of the EHM. Since we have applied the same
mapping procedure and got a similar low-energy field theory, the different phases are 
characterized by the same field structure as in Ref.~\cite{nakamura2000}. 
For $V\lesssim U/2.18$, the system is in a SDW phase,
with a gapless spin sector and dominant correlations of the form
$\langle (n_{\uparrow,i}-n_{\downarrow,i})(n_{\uparrow,i+x}-n_{\downarrow,i+x})\rangle$. On the other hand, a (mass) density-wave is formed
above the critical value $V>V_c^{(G)}$, where the dominant correlations
are of the form $\langle
 (n_{\uparrow,i}+n_{\downarrow,i})(n_{\uparrow,i+x}+n_{\downarrow,i+x}) \rangle$. In the intermediate regime 
$V_c^{(BKT)}<V<V_c^{(G)}$, neither SDW nor CDW order are stable, 
and the system exhibits a BOW, characterized by both a 
finite spin and density gap, and a dominant order described by the parameter
\begin{eqnarray}\label{eq:Bi}
\langle B_i\rangle =\langle\frac{1}{2}\sum_\sigma\left(c^\dag_{\sigma,i}c_{\sigma,i+1}+c^\dag_{\sigma,i+1}c_{\sigma,i}\right)\rangle.
\end{eqnarray}
This indicates that dimers are spontaneously formed on nearest-neighbor bonds, a phenomenon usually called spontaneous 
spin-Peierls dimerization. In analogy with the EHM analysis, the corresponding charge field is pinned at the value $\langle\varphi_c^*\rangle=
0$, contrary to the SDW phase, where $\langle\varphi_c^*\rangle\neq0$. This behavior is triggered by the pre factor of the mass gap, $g_c$: according to its sign, the 
cosine potential is either pinned around $0$ or $\pi/2$, which implies different expectation values for $\varphi_c$.

\begin{figure}[t]
\begin{center}
\includegraphics[height=65mm]{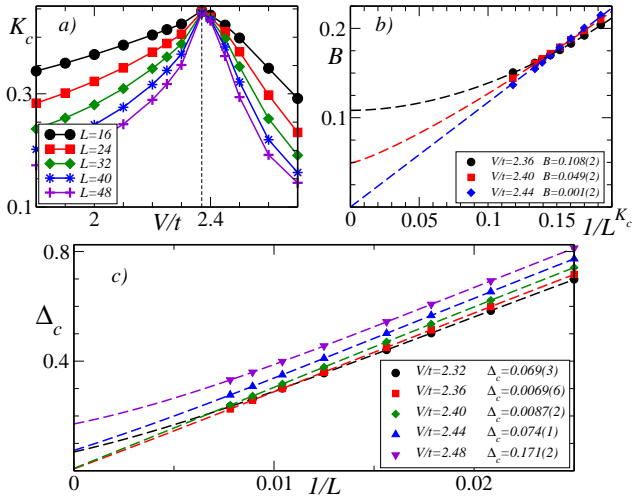}
\caption{(Color online) (a):  Luttinger parameter $K_c$, for $U=4t$ as a function of $V/t$ for different system sizes $L$. (b): BOW order parameter $\langle B\rangle$ for various $V/t$, in the box is its value in the TDL. (c): Density gap $\Delta_c$ for various values of $V/t$, in the box its values in the TDL. Straight lines are guides for the eye, whereas dashed lines fit the numerical data using the appropriate scaling laws.}\label{fig2}
\end{center}
\end{figure}

As such, we conclude that, once  the estimates of the sine-Gordon model parameters from $H_{12}$ are combined with a non-perturbative
treatment of the single species interactions, the low-energy
theory already predicts a finite region of parameter space where bond
order is stable. This is due to the effects of the long-range dipolar tails, 
which consistently affect the Luttinger parameter of the single species
Hamiltonians, and lead to different conditions for the BKT and Gaussian
transitions.

However, the treatment is supposed to work only in the weak-coupling
regime since on-site interactions are taken into account only perturbatively, 
and the provided estimate on the Gaussian line does not capture indeed the entire dipolar interaction. 
Moreover, treating the interchain interactions non-perturbatively requires the SU(2) symmetry to be 
reinforced at low-energies as discussed above; the quantitative accuracy of this procedure is 
hard to establish {\it a priori}. 
In order to confirm the existence and the finite extent of a BOW phase at finite, intermediate couplings, 
a non-perturbative approach is required. In the next section, we provide a numerical analysis of Eq.~\eqref{H_dis}
using the present theory as a guideline to identify the possible transitions in the microscopic model.

\section{Numerical results}

The accurate determination of the phase diagram of the system is a very challenging task. As far as the simpler EHM is concerned, despite the great effort put in its numerical study over the last decade, few results have shown  good agreement with each other~\cite{nakamura2000,sengupta2002,furusaki2002,
jeckelmann2002,sandvik2004,furusaki2004,tam2006,ejima2007}. Mindful of the difficulties in the determination of the phase diagram for such a system, we carefully calibrated on the EHM the methods employed below to detect phase transitions
with both periodic (PBC) and open boundary conditions (OBC); discarded weights of the order $10^{-8}$ allow us to estimate the phase transitions within $1\%$ with the most recent and accurate results\cite{sandvik2004,ejima2007}.

We calculate several physical quantities, summarized in Table \ref{table1}, to determine as accurately as possible the two phase boundaries of Eq.~\eqref{H_dis}, keeping in the DMRG simulation terms up to $|j-i|=5$ in the off-site dipolar interaction part. We verified for small system sizes that the inclusion of longer range terms does not  significantly affect the quantities of interest.

\begin{figure}[t]
\begin{center}
\includegraphics[height=60mm]{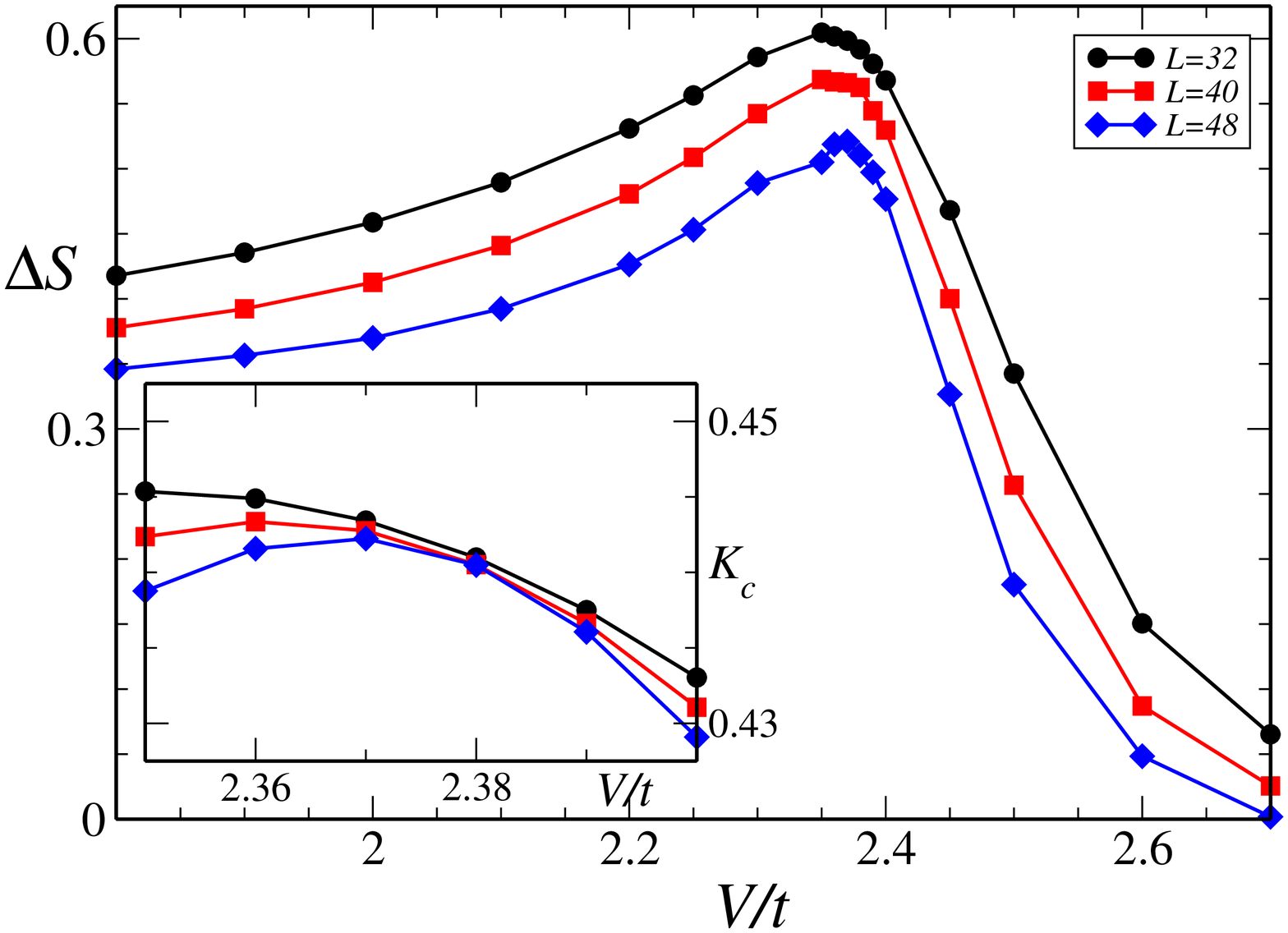}
\caption{(Color online) The difference $\Delta S(L)=S_L(L/2)-S_{L/2}(L/4)$ for several system sizes for $U=4t$ varying $V/t$.  The inset shows the Luttinger parameter $K_c$ in the region $2.35\leq V/t\leq2.4$. Straight lines are guides for the eye.}\label{fig3}
\end{center}
\end{figure}

\subsection{CDW-BOW transition} 
Starting in the CDW phase, as the dipolar interaction strength is decreased the system enters in a BOW phase  through a Gaussian phase transition (away from the strong coupling limit). As a first probe of the transition, we calculate the density gap
\begin{equation}
\Delta_c=\lim_{L\rightarrow\infty}\left[E(N+2,0)+E(N-2,0)-2E(N,0)\right]/2
\end{equation}
where $E(N,S_z)$ is the ground state energy of $N=L$ particles with total magnetization $S_z=(N_\uparrow-N_\downarrow)/2$. Due to the competition between long range and on-site interactions the finite-size gap has to take a minimum value at the transition point (which is gapless in the thermodynamic limit~\cite{ejima2007}). To locate the phase boundary as precisely as possible we also calculate the BOW order parameter defined as 
\begin{equation}
\langle B\rangle=\lim_{L\rightarrow\infty}|\langle B_{L/2}-B_{L/2+1}\rangle|,
\end{equation}
where $B_i$ is the operator defined in Eq.~\eqref{eq:Bi}. $\langle B \rangle$ is the amplitude of the oscillation of the BOW operator at the center of the chain, defined in such a way that Friedel oscillations are weaker~\cite{ejima2007}: a non-vanishing value of $\langle B\rangle$ will be a clear signature of the BOW phase.

\begin{table}[t]
\caption{Thermodynamic properties of the different phases discussed in the main text (table on top) and behavior of the various observables at the transition lines (bottom table).}\label{table1}
\vspace{0.1cm}
\centering
\begin{tabular}{|c |c|c|c|}
\hline
\hline
& \textit{CDW} & \textit{BOW} & \textit{SDW} \\
\hline\hline
$\Delta_c$&$>0$& $>0$ &  $>0$\\
\hline
$\Delta_s$ & $>0$ & $>0$ & 0\\
\hline
$\langle B\rangle$& 0& $\neq0$ & 0\\
\hline\hline
\end{tabular}

\vspace{.35cm}

\centering
\begin{tabular}{|c|c|}
\hline
\hline
\textit{CDW-BOW} & \textit{BOW-SDW} \\
\hline\hline
$\Delta_c=0,\Delta_s>0$ &  $\Delta_s=0, \Delta_c>0$\\
\hline
$K_c\neq 0$ & $K_s=1$\\
\hline
$\langle B\rangle=0$& $\langle B\rangle=0$\\
\hline\hline
\end{tabular}
\end{table}

A further signature of the phase transition is given by the Luttinger liquid (LL) parameter $K_c$. For a periodic chain it can be extracted from the (density) static structure factor
\begin{eqnarray}\label{csf}
S_c(q)=\frac{1}{L}\sum_{k,l}e^{iq(k-l)}\left(\langle n_k n_l\rangle\ -\langle n_k\rangle\langle n_l\rangle\right)
\end{eqnarray}
with $q=2\pi/L$. Within LL theory $K_c=\lim_{q\rightarrow 0}\pi S_c(q)/q$ is finite only on the phase transition line for a continuous transition (while it is always zero instead for first order transitions). 
Since we are dealing with a finite size system we expect to see a sharp peak at the transition line~\cite{sandvik2004}.

In order to get the correct values in the thermodynamic limit for the quantities described above, 
a careful finite size scaling analysis must be carried out. The density gap $\Delta_c$ is extrapolated fitting the data with a fourth-order polynomial in $1/L$, reproducing the holon band structure near the band edges.
For the BOW order parameter $\langle B\rangle$ we assume that in the center of the chain the amplitude of Friedel oscillations is proportional to $1/L^{K_c}$ \cite{ejima2007} to extrapolate its thermodynamic value.

As an additional benchmark to pinpoint the Gaussian transition point, we study the behavior of 
the von Neumann entropy, that can be successfully used to locate critical points~\cite{kollath}.
In particular, we monitor the following quantity
\begin{eqnarray}\label{eq_ent}
\Delta S(L)=S_L(L/2)-S_{L/2}(L/4)
\end{eqnarray}
i.e., the increase of the entropy at the mid-system interface upon doubling the system size; $S_L(l)$ denotes the von Neumann entropy of a block of size $l$, $L$ being the length of the whole system. To avoid boundary terms that may give rise to oscillating corrections to the entropy, we will impose PBC on the system under study: due to the finite size of the system we expect $\Delta S(L)$ to develop a peak at the critical point.

We run DMRG simulations for several values of the on-site interaction $U$ varying the strength of the dipolar interaction $V$. To calculate the density gap and the BOW order parameter we employ OBC varying the  system sizes  between $L=32$ and  $L=128$. We keep up to $m=1256$ states and perform six sweeps in the renormalization procedure: the corresponding truncation errors are at most of order $10^{-8}$. To calculate the static structure factor (\ref{csf}) and the increase of the entropy (\ref{eq_ent}), we instead use PBC on smaller systems ($L=16,20,24,32,40,48$), keeping up to $m=1400$ states and six sweeps.

\begin{figure}[t]
\begin{center}
\includegraphics[height=65mm]{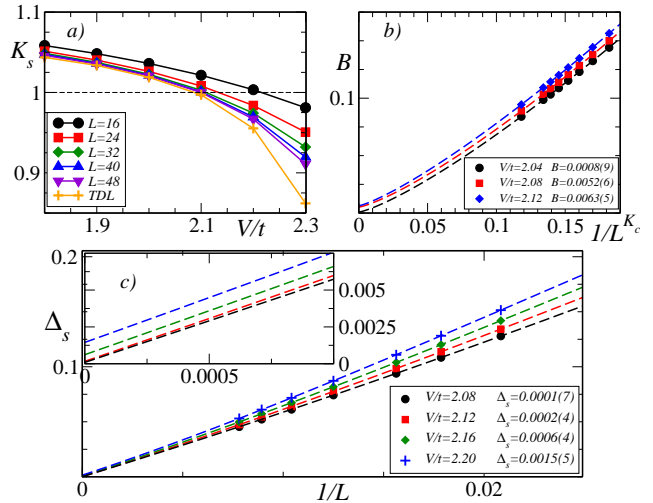}
\caption{(Color online) (a):  Luttinger parameter $K_s$, for $U=4t$ as a function of $V/t$ for different system sizes $L$. (b): BOW order parameter $\langle B\rangle$ for various $V/t$, in the box is its value in the TDL. (c): Spin gap  $\Delta_s$ for various values of $V/t$, in the box its values in the TDL; the inset shows the region around the origin. Straight lines are guides for the eye, whereas dashed lines fit the numerical data using the appropriate scaling laws.}\label{fig4}
\end{center}
\end{figure}

As a case study we discuss in detail the results for $U=4t$, the same analysis having been carried on for all the points depicted in the phase diagram Fig.~\ref{fig1}. In  panel (a) of Fig.~\ref{fig2} and in the inset of Fig.~\ref{fig3}  one sees that  $K_c$ develops a peak (sharper as the system size increases) when $V_c/t\simeq2.37$, that is also the same  $V_c/t$ where the peak in $\Delta S$ is located as shown in Fig.~\ref{fig3}. In panel (b) of Fig.~\ref{fig2} the BOW order parameter $\langle B\rangle$ is plotted as a function of $1/L^{K_c^\ast}$, with $K^\ast_c\simeq0.44$ the thermodynamic limit (TDL) value of $K_c$ close to the transition line: an accurate finite size scaling shows that $\langle B\rangle$ vanishes as $V/t=2.44$ while it is still finite for $V/t=2.4$. Finally, in panel (c) of Fig.~\ref{fig2} we plot the density gap $\Delta_c$, and after a careful extrapolation in the TDL  we see that $\Delta_c$ vanishes as $V/t=2.36(4)$.   
The results discussed above are all in quantitative agreement, leading us to infer that the transition between the CDW and the BOW phases occurs for $V_c/t=2.40(4)$.

\begin{figure}[t]
\begin{center}
\includegraphics[height=65mm]{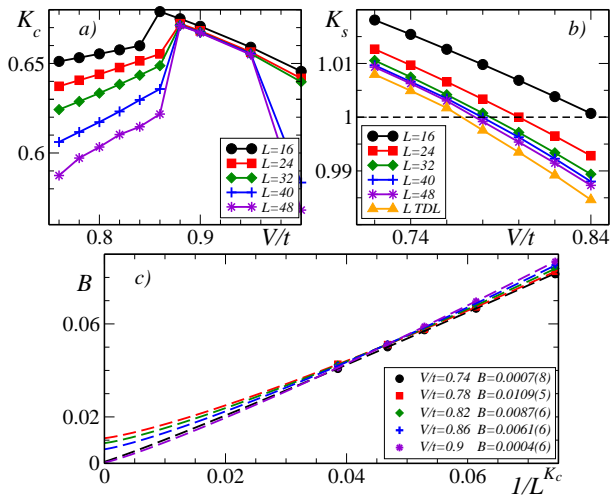}
\caption{(Color online) (a): Luttinger parameter $K_c$ for $U=1.5t$ as a function of $V/t$: the maximum location indicates the Gaussian transition point, $V_c/t = 0.88\pm0.04$. (b): Luttinger parameter $K_s$ for $U=1.5t$ as a function of $V/t$ indicating a BKT transition (see text). (c): bond order parameter finite-size scaling at $U=1.5t$ and different values of the dipolar interaction strength. }\label{fig5}
\end{center}
\end{figure}

\subsection{SDW-BOW transition} As the strength of the dipolar interaction is further decreased, Peierls dimerization is destroyed and the system enters in a SDW with a uniform distribution of the density and no gap in the spin sector. The BKT nature of this transition makes its location challenging when evaluating the spin gap
\begin{equation}
\Delta_s=\lim_{L\rightarrow\infty}\left[E(N,1)-E(N,0)\right]
\end{equation}
since it is exponentially small close to the transition line\cite{bosonization}. A valid alternative is provided by the
spin-static structure factor
\begin{eqnarray}
S_s(q)=\frac{1}{L}\sum_{k,l}e^{iq(k-l)}\left(\langle s_k^zs_l^z\rangle\ -\langle s_k^z\rangle\langle s_l^z\rangle\right).
\end{eqnarray}
Indeed LL theory predicts that $K_s=\lim_{q\rightarrow 0}\pi S_s(q)/q$ is zero in the spin gapped phase and $K_s=1$ in the gapless one. 
Logarithmic corrections prevent $K_s$ to reach the latter value even for long chains. 
Nonetheless, such corrections have been shown to vanish in the frustrated $J$-$J^\prime$ model when the system forms dimers\cite{eggert}. 
The same is true for the EHM\cite{sandvik2004} since the BOW-SDW transition should have the same nature, and it is expected to hold also in our case. 
In the BOW phase $K_s=0$ near the transition only for very large systems. Following \cite{sandvik2004}, we estimate the transition point when, at fixed $U$, $K_s$ crosses 1 from above as $V$ is increased. The thermodynamic limit of the spin gap $\Delta_s$ is obtained, as for the density gap, fitting the data with a fourth-order polynomial in $1/L$, reproducing the spinon band structure near the band edges. 

The results of the DMRG simulations are reported in Fig.~\ref{fig4}. First of all one can see in panel (a) of Fig.~\ref{fig4} that $K_s$ crosses 1 for $V/t\le 2.1$ and an accurate finite size scaling allows us to locate the transition point for $V/t=2.089$. This result is confirmed looking at the BOW order parameter $\langle B \rangle$ in panel (b) of Fig.~\ref{fig4}: in the thermodynamic limit $\langle B \rangle$ is still finite at $V/t=2.08$ while it vanishes for smaller values of the strength of the dipolar interaction. We therefore conclude that the transition between the SDW and the BOW phases is located at $V_c/t=2.08(4)$. For the sake of completeness, even if it is not conclusive about a precise determination of the critical point, we plot in  panel (c) of Fig.~\ref{fig4} the spin gaps: the extrapolated values close to the critical point are too small to be resolved, as expected from the BKT nature of the transition. As such, a reliable estimate of the transition point can hardly be drawn from this quantity alone. 

A second set of results is illustrated at the smallest coupling we have analyzed, $U=1.5t$, in Fig.~\ref{fig5}. There, the transition point estimates extrapolated from the Luttinger parameter still point towards a finite extent of the BOW region for $0.77(3)< V/t<0.88 (4)$. However, the magnitude of the bond order parameter is notably reduced with respect to the $U=4t$ case.

\section{Experimental regimes and probes}
Polar molecules and magnetic atoms offer strong versatility in tuning interactions. 
In the latter case, the ratio $V/U$ can be independently tuned by means of 
Feshbach resonances, which have been already reported for bosonic 
$^{52}$Cr~\cite{LewensteinReview} and $^{167}$Er~\cite{aikawa2012}
isotopes. In the case of molecular gases stable under two-body recombination, 
accurate estimates of on-site interactions will be required. An alternative
approach can employ Feshbach molecules of strongly magnetic atoms such 
as Er or Dy, effectively increasing by a factor $\sim8$ the strength of the dipole-dipole
interactions~\cite{dalmonte2010}.

\begin{figure}[t]
\begin{center}
\includegraphics[height=58mm]{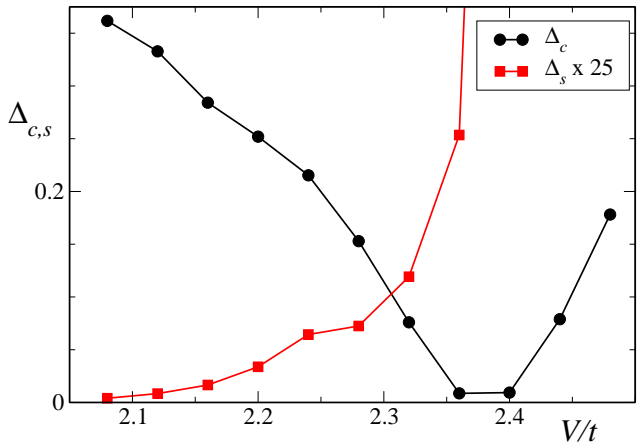}
\caption{(Color online) Density and spin ($\times25$) gaps in the thermodynamic limit for various $V/t$.}\label{fig6}
\end{center}
\end{figure}

The existence of the BOW phase can be indirectly probed spectroscopically
as follows: first, the density gap is estimated by means of lattice modulation
spectroscopy, indicating the onset of the Gaussian transition. Subsequently, 
the spin gap is estimated by means of RF spectroscopy.
The BOW phase can be located in the intermediate region between the 
two transition points, as illustrated in Fig.~\ref{fig6}. A more solid, direct probe
of the exotic nature of the insulating state is the long-range nature of 
the parity order parameters:
\begin{equation}
O_s(x)=\langle \prod_{j=\ell}^{\ell+x}e^{i\pi S^z_{j}}\rangle, 
\quad O_c(x)=\langle \prod_{j=\ell}^{\ell+x}e^{i\pi n_{j}}\rangle, 
\end{equation}
which are well-defined order parameters for general Hubbard models~\cite{montorsi2012}.
As discussed in Ref.~\cite{montorsi2012,barbiero2013} such correlation functions properly distinguish the 
CDW, SDW and BOW phases. In particular only the latter phase has long-range order in both $O_s$ and $O_c$.
Although measuring parity order parameters is a difficult task, recently $O_c$ has been measured for cold atoms with short range interaction loaded in an optical lattice by {\it in situ} imaging of the many-body wave function via atom fluorescence~\cite{endres2011}.

As far as thermal effects are concerned, the major experimental challenge would be to reach regimes where the temperature is smaller than the finite-size spin gap, which is at most of order $0.15t$ for relatively small systems deep in the BOW phase. This implies that temperatures of order of $10$nK would be required in order to neglect thermal effects. While this is indeed a challenging task, we notice that recent experiments using both Cr atoms \cite{depaz2013} and RbK polar molecules  \cite{yan2013} have demonstrated coherent dipolar spin dynamics in the 50 Hz range, in the similar regime of the aforementioned temperatures.

\section{Conclusions}
In summary, we have provided a detailed study of how dipolar fermionic mixtures support exotic
insulating states with dominant bond-order. We have underpinned the 
corresponding Hubbard phase diagram by combining numerical
and analytical techniques. Remarkably, differently from the standard EHM, we find that the BOW phase can be  found already at the level of one-loop \textit{g}-ology. This is due to the long-range nature of the interaction which enhances frustration effects in the system, shifting as well the BOW region to larger values of the off-site interaction $V$. While an accurate determination of the spin gap is very challenging, precise estimates of the phase boundaries of the BOW region can be given via finite-size-scaling of the Luttinger parameters, which, as in the case of the EHM~\cite{sandvik2004, ejima2007}, represent a very efficient and precise method to underpin BKT transitions. 
Experimentally, the non-trivial correlations embodied in the
bond-density-wave phase can be faithfully captured by string order parameters by means of {\it in situ} imaging.  This makes cold atoms in optical lattices
an ideal setup for the investigation and demonstration 
of bond-order instabilities in strongly correlated systems. 

\acknowledgments
We thank S. Bhongale, E. Ercolessi, L. Mathey, A. Montorsi, G. Pupillo and M. Roncaglia for fruitful discussions, and F. Ortolani for help with the DMRG code. This work is supported by AQUTE, the Austrian Science Fund FWF (SFB FOQUS F4015-N16), ERC (QGBE) grant, ERC (UQUAM) grant, Provincia Autonoma di Trento, Cariparo Foundation  (Eccellenza Grant 11/12) and ANR 2010 BLANC 0406-0. L. B. is grateful to HPC resources of CALMIP (Toulouse) and CNR-INO BEC Center (Trento) for CPU time.

\end{document}